# Weak Anchoring and Surface Elasticity Effects in Electroosmotic Flow of Nematic Liquid Crystals Through Narrow Confinements

Antarip Poddar, Jayabrata Dhar and Suman Chakraborty
Department of Mechanical Engineering,
Indian Institute of Technology Kharagpur, Kharagpur-721302, India

**Abstract**

Advent of nematic liquid crystals flows have attracted renewed attention in view of microfluidic transport phenomena. Among various transport processes, electroosmosis stands as one of the efficient flow actuation method through narrow confinement. In the present study, we explore the electrically actuated flow of a nematic fluid with ionic inclusions taking into account the influences from surface induced elastic and electrical double layer phenomena. Influence of surface effects on the flow characteristics is known to get augmented in micro-confined environment and must be properly addressed. Towards this, we devise the coupled flow governing equations from fundamental free energy analysis considering the contributions from first and second-order elastic, dielectric, flexoelectric, ionic and entropic energies. We have further considered weak anchoring surface conditions with second order elasticity which helps us to more accurately capture the director deformations along the boundaries. The present study focuses on the influence of surface charge and elasticity effects in the resulting linear electroosmosis through a slit-type microchannel whose surface are considered to be chemically treated in order to display a homeotropic-type weak anchoring state. An optical periodic stripe configuration of the nematic director has been observed especially for higher electric fields wherein the Ericksen number for the dynamic study is restricted to the order of unity.

___________________________
email for correspondence: suman@mech.iitkgp.ernet.in

**Introduction**

Electrokientic transport phenomena of complex fluids through micro-condiment have been elaborately studied in the literature [1–3] due to its various applications in biomedical engineering [4,5], energy conversion processes [6,7], environmental sciences and thermal management of electronic packages, to name a few. Emergence of electrokinetic transport of ordered fluids, especially of anisotropic liquid crystal medium, has led to numerous studies in recent times that explores the flow behavior and non-linear effects under the scope of micro-scale dynamics [8]. Nematic Liquid Crystals (NLCs) are among such ordered fluids that display an orientational order across the study length scales [9,10]. The molecules of NLCs have, in general, rod-shaped structures and remains arranged with a typical specific order. The average molecular long-axis alignment of such NLC molecules is denoted using a unit vector **n**, known as the director [9,10]. When confined within a microchannel, NLCs show intriguing elastic and flow characteristics to external stimuli, which have recently motivated numerous microfluidic studies of such nematic cells [11–14]. In the context of flow actuation through narrow conduits, electroosmosis, defined as the mechanism of actuating a fluid in contact with a charged surface by the application of an external electric field [15,16], has emerged as a promising means of energy efficient flow actuation process. Such flows are generally characterized by a charged fluid layer adjacent to the surface known as the electrical double layer or EDL that gets induced due to certain physico-chemical interactions. A balance between the electrostatic and the thermal interactions among the ionic species result in a charge distribution across the channel with a dominant counterion, ions of opposite polarity to that of the substrate, presence at the vicinity of the surface. Upon application of a longitudinal electric field, an advection of the surplus ions within the EDL sets in, which, as a consequence of viscous effects, drags the solvent molecules along with them resulting in electroosmotic flow. Such linear electroosmotic flows, wherein the flow velocity linearly depends with the applied field, have been rigorously studied in electrokinetic literature [15–18]. Later on, non-linear electroosmosis around polarizable surfaces such as metallic colloids, where the slip velocity varies quadratically with the applied electric field [19–21], is discovered within the purview of Induced Charge Electroosmosis (ICEO). Very recently Lazo and co-workers [8] experimentally demonstrated a non-linear electroosmotic phenomenon in nematic liquid crystals exploiting the spatial charge separation owing to the anisotropy in electrical conductivity and consequent director distortion. They showed that although the presence of ionic currents in LCs has traditionally been considered undesirable in display related applications, liquid crystal enabled electroosmosis phenomenon can be turned into a great advantage if efficiently used as a non-mechanical fluidic transport technique in microfluidic applications. However, sustained flow actuation employing a DC field in the scope of linear electroosmosis through a narrow conduit has never been studied for the case of complex NLC liquids.

A critically prominent factor for NLC dynamics within a narrow confinement is that the macroscopic behavior of NLC director greatly depends on its interaction with its confining solid substrates besides other external factors [10,11]. Owing to high surface to volume ratio of nematic cells, the boundary effects propagate far into the bulk nematic medium, and consequently, pose a significant influence on the equilibrium director distortion and velocity distribution [11,22–24]. In the absence of any external perturbations, the NLC director gets oriented in a certain preferential direction at the substrate-fluid interface, denoted as its easy direction. Upon application of an external field the orientation of liquid crystal molecules at the interface may deviate from the easy direction giving rise to a phenomenon known as 'weak anchoring'. Such forms of weak surface alignment of directors has been realized in various experimental studies which include soft rubbing of a polymer film, oblique evaporation of $SiO_2$ [25], photo-induced ordering [26] and chemical patterning of surfaces [27,28]. Along with such experiments, parallel theories have also been developed to account for the surface-induced influence on the resulting director orientation. Rapini and Papoular [29] proposed that the weak anchoring condition stems from an additional preliminary surface energy contribution to the total free energy of the nematic cell. Coupling the weak anchoring condition with the first-order elastic theory for the dynamics of NLCs, introduced by Frank and Oseen [10], satisfactorily captures the director field in the bulk, however, it fails to explain the strong director deformation observed close to the nematic-substrate interface. Later on, it was showed [30,31] that a new energy term containing the surface-like elasticity due to mixed splay-bend contribution (with elastic constant $K_{13}$) must be added to the free energy contribution. However, such modification to the elastic free energy makes it unbounded from below resulting in discontinuity in the director orientation at the surface [32–34]. This paradox is resolved afterwards by including higher order elasticity terms in the bulk free energy contribution [35,36]. The sharp director variation, that is observed in an extremely thin transition region near the surface having a characteristic length scale of the order of molecular interaction [35,37], may then be successfully captured by considering a more accurate second-order elastic theory pertinent in this narrow sub-layer [38–40].

In the present study, we address the sustained flow actuation mechanism within a NLC cell employing linear electroosmosis. NLCs have been shown to induce electrical charged layers adjacent to the substrate due to surface charge adsorption in the presence of ionic impurities within the NLC medium [41–44]. We have relaxed the point-charge approximation for the ionic impurities to include the excluded-volume effects of the finite hydration shell size [45]. We generalize the study considering a weak anchoring boundary condition of the director at the fluid-substrate interface. In order to comprehend the underlying physics and accurately capture the near-surface effects within narrow confinement, a second-order elastic energy besides the classical first-order Frank-Oseen elastic energy have been incorporated to study the NLC director deformation. It is noteworthy that, besides the director elastic energies, dielectric and flexoelectric energies due to the presence of an electric field as well as energies from the contribution of the ionic species distribution must be carefully taken into account in order to model the flow of NLC fluids. For the nematodynamic estimation, we employ the classical Leslie-Ericksen flow model [10] for the NLC governing the anisotropic fluid flow characteristics. Here we focus on the effect of second-order surface elasticity, surface weak anchoring energy and induced surface charge density on the electroosmotic flow considering a homeotropic easy direction arrangement of the NLC director at the surface boundaries. Formation of optical periodic stripes of the director configuration is observed especially for higher electric fields. Such observations are common for director arrangements in NLC flows with surface confinements [12,22].

## Mathematical Formulation

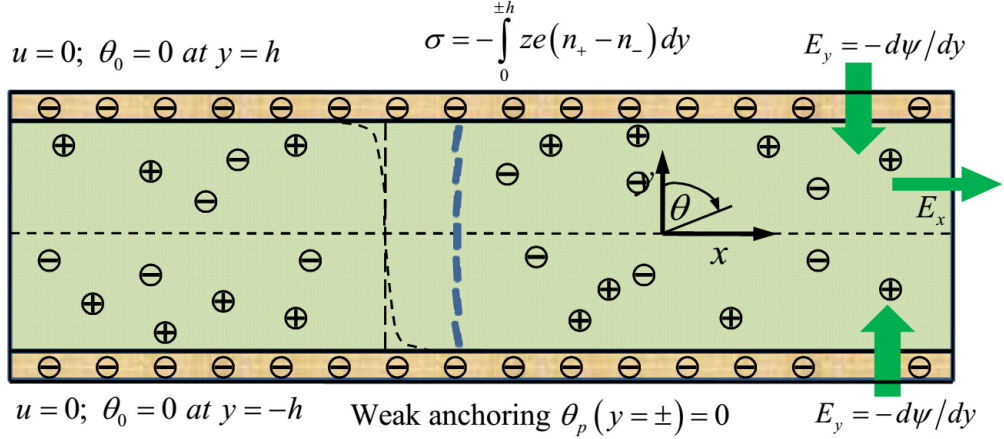

**Fig 1:** Schematics representing the electroosmosis of an anisotropic nematic liquid under the action of an external axial electric field. An Electrical Double Layer get induced adjacent to each of the substrate interface that, besides influencing the fluid rheology and director orientation, provides the genesis of the flow actuation body force. A weak boundary condition with homeotropic-type easy axis is considered at both the surfaces.

For the present study, we consider a NLC with splay and bend elastic coefficients as $K_{11}$ and $K_{33}$, respectively, confined between two semi-infinite parallel wall having a separation of *2h* as shown in the schematics (figure 1). The nematic liquid is associated with an intrinsic dielectric anisotropy due to its distinct parallel and perpendicular dielectric constant represented by $\varepsilon_{\parallel}$ and $\varepsilon_{\perp}$, respectively, while its flexoelectric coefficient is given using $e_1$ and $e_3$. The average direction of the nematic molecules, represented by unit director vector **n**, is assumed to vary across the channel width (*y*-axis) with the restriction of planar deformation (director deformations remain in the flow plane). Consequently the unit director may be reformulated in the form $\mathbf{n} = \sin\theta(y)\mathbf{i} + \cos\theta(y)\mathbf{j}$ as shown in the above schematic while a weak anchoring condition of the director prevails at both the walls. Weak anchoring refers to the condition wherein the director orientation at any surface interface is evaluated by the balance of relevant surface energies. Furthermore, in our study, we consider the existence of an induced EDL at the liquid-substrate interface due to certain physico-chemical interactions which impose a non-linear distribution of the charged entities dissolved in the liquid medium having a number density $n_0$ in the reservoir [41,46,47]. Owing to this charge at the interface, with a surface charge density $\sigma_w$, and the ionic charge distribution in the liquid domain, a transverse non-uniform electric field $E_y(y)$ gets spontaneously induced which, besides affecting the fluid rheology and the anchoring conditions, provides the necessary body force for the flow actuation of the liquid medium. Upon the application of an external longitudinal electric field $E_x$ electroosmotic flow results. Here the axial velocity field is assumed to be only a function of the transverse direction $\mathbf{V} = u(y)\mathbf{i}$. An interesting aspect of electro-nematodynamic flows with weak anchoring is the coupling interplay among the director orientation, potential distribution and flow velocity resulting in an intriguing non-Newtonian flow characteristics, and therefore, a careful analysis of the underlying physics through the total free energy minimization formulation must be performed. We have further assumed a negligible conductive anisotropy for the NLC with a consequence that the charge separation due to the Carr-Helfrich effect [48] remains absent. This assumption, consistent with numerous previous studies [49–52] which explicitly considered the ionic presence within the LC phase but neglected charge separation phenomena, allows us to focus the present study within the domain of linear electroosmosis. Following the aforementioned notion, we proceed to evaluate the potential distribution within the EDL and equilibrium director configuration of the nematic phase with weak anchoring conditions by considering the total free energy $F$ that incorporates the elastic energy of the nematic molecules, the dielectric energy inherent to its anisotropic nature, the flexoelectric energy attributed to shape-induced polarization of the molecules, the internal energy and the entropic contributions that accounts for the ionic charge distribution and their excluded volume effects.

The elastic energy associated with the director deformation in the NLC phase reads [53–56]

$$F_{elast} = \int_V f_V dV + \int_S (f_{13} + f_{24} + f_S) dS \quad (1)$$

where the integrals are taken over volume $V$ and surface $S$ of the nematic sample. The bulk free energy $f_V$ due elastic distortion of the NLC director is obtained from the first order elasticity theory as proposed by Frank and Oseen [9,10] which takes the form

$$f_{V,\text{1st order}} = \frac{1}{2}\left(K_{11}(\nabla\cdot\mathbf{n})^2 + K_{22}(\mathbf{n}\cdot\nabla\times\mathbf{n})^2 + K_{33}(\mathbf{n}\times\nabla\times\mathbf{n})^2\right) \qquad (2)$$

In order to capture the sharp variation of the director field within the surface transition layer and frame a well posed variational problem [57], we resort to the 'second order elastic theory' [36,55]. The general expression of second-order free energy density is rather complex involving a set of 35 new elastic constants. This makes it practically impossible to solve for the equilibrium director field. However, it was later found [36] that close to a threshold where the distortion amplitude is very small, the additional term that remains is given by $K^*\left(\nabla^2\mathbf{n}\right)^2 \approx K^*\left(\dfrac{d^2\theta}{dy^2}\right)^2$, where $K^*$ is the bulk second order elastic constant. Thus, in this limiting case the resultant bulk free energy density is defined by [32,33,36,55]

$$f_{V,\text{2nd order}} = \frac{1}{2}\left(K_{11}(\nabla\cdot\mathbf{n})^2 + K_{22}(\mathbf{n}\cdot\nabla\times\mathbf{n})^2 + K_{33}(\mathbf{n}\times\nabla\times\mathbf{n})^2\right) + K^*\left(\nabla^2\mathbf{n}\right)^2 \qquad (3)$$

Though the consideration of a bulk second order elastic constant is valid in the surface transition layer, it fails to track the larger distortion of directors far inside the nematic cell. Hence we employ a two layer model [38] where the flow domain is divided into a sub-surface region spanning upto a distance $\delta$ in the vicinity of each wall and a bulk-layer covering rest of the nematic cell. The transition layer thickness $\delta$ is considered to be of few characteristic length $\left(\sqrt{2K^*/K_{11}}\right)$ [38]. Towards this, we employ the first and second order elastic theories in the bulk and surface layer, respectively. Here $f_{13}$ and $f_{24}$ describes the second-order surface elastic energy terms given by $f_{13} = K_{13}\upsilon\cdot\mathbf{n}(\nabla\cdot\mathbf{n})$ and $f_{24} = -\dfrac{1}{2}(K_{22}+K_{24})\upsilon\cdot\left[\mathbf{n}(\nabla\cdot\mathbf{n}) + \mathbf{n}\times\nabla\times\mathbf{n}\right]$, respectively; $\upsilon$ represents the unit surface normal; $K_{13}$ denotes the mixed splay-bend elastic constant and $K_{24}$ denotes the saddle-bend elastic constant. Further $f_S$ stands for free energy density of nematics-substrate interaction given by [29,39]

$$f_S = \frac{1}{2}W_S\sin^2(\theta-\theta_p) \qquad (4)$$

where $W_S$ and $\theta_p$ are the anchoring energy constant and the director orientation along the easy axis, also known as the pretilt angle, respectively.

Besides, the elastic energy due to director deformation, an additional energy in presence of an electric field $F_{el}$ gets associated with the NLC phase having dissolved ionic impurities. The cumulative electrical energy incorporates the energies originating from liquid dielectric anisotropy ($F_{de}$), gradient flexoelectric molecular nature $F_{fe}$ and internal energy $F_{int}$ due to the presence of free ions as $F_{el} = F_{de} + F_{flex} + F_{int}$. The dielectric energy density may be evaluated from the classical description $F_{de} = -\frac{1}{2}\int \mathbf{D} \cdot \mathbf{E} dV$ where the electric displacement vector $\mathbf{D}$ defined in case of liquid crystals as $\varepsilon_0 \left[ \varepsilon_\perp \mathbf{E} + \varepsilon_a (\mathbf{E} \cdot \mathbf{n}) \mathbf{n} \right]$ and the electric field vector is given by $\mathbf{E} = E_x \mathbf{i} + E_y(y) \mathbf{j}$. Here $\varepsilon_a = \varepsilon_\parallel - \varepsilon_\perp$ is known as the dielectric anisotropy, $\varepsilon_0$ is the absolute permittivity of free space, $E_x$ denotes the applied axial field, $E_y = -\nabla \psi(y)$ denotes the spontaneously induced inhomogeneous transverse field and $\psi(y)$ represents the potential distribution within the EDL. This results in the anisotropic dielectric energy of the form

$$F_{de} = \int -\frac{\varepsilon_0 \varepsilon_a}{2} \left[ E_x \sin(\theta) + E_y(y) \cos(\theta) \right]^2 - \frac{\varepsilon_0 \varepsilon_\perp}{2} \left[ E_x^2 + E_y(y)^2 \right] dy \tag{5}$$

The flexoelectric counterpart of the energy density for the nematic molecules is determined using $F_{flex} = -\int \mathbf{P}_{fl} \cdot \mathbf{E} dV$ where the induced polarization for such ordered nematic is given by [58] $\mathbf{P}_{fl} = e_1 (\mathbf{n} \nabla \cdot \mathbf{n}) + e_3 (\mathbf{n} \times \nabla \times \mathbf{n})$, where $e_1$ and $e_3$ are the flexoelectric coefficients. The resultant flexoelectric energy functional reads

$$F_{flex} = \int \left( \left[ \left( e_1 \sin^2(\theta) + e_3 \cos^2(\theta) \right) E_x + (e_1 - e_3) \sin(\theta) \cos(\theta) E_y(y) \right] \frac{d\theta}{dy} \right) dy \tag{6}$$

The contribution to the internal energy is from the dissolved ionic species within the nematic sample, which comprises of the ionic electrostatic energy having the form [59]

$$F_{int} = \int ze\phi(x, y)(n_+ - n_-) dV \tag{7}$$

where the associated total potential due to the combined applied and induced electric field reads $\phi(x, y) = \psi(y) + (\phi_0 - x E_1)$.

It must be appreciated that the dissolved ions usually have finite size effects that should be taken into consideration which restricts excessive ionic crowding near the wall particularly for situations involving high ionic concentration and strong electrostatic interactions. Relaxing the point charge approximation, the entropic contribution considering the finite ionic shell size is given by the form [60,61]

$$F_{entropic} = -TS = k_B T \int dy \left[ n_+ \ln(a_+^3 n_+) + n_- \ln(a_-^3 n_-) - n_+ - n_- \right] \\ + \frac{k_B T}{a^3} \int dy \left[ (1 - a_+^3 n_+ - a_-^3 n_-) \ln(1 - a_+^3 n_+ - a_-^3 n_-) \right] \tag{8}$$

The above formulation allows the inclusion of excluded volume effects within the continuum modeling of the ionic distribution where the number density of positive (negative) ions is given by $n_+ (n_-)$ while their corresponding ionic shell size is denoted using $a_+ (a_-)$. Further for the sake of simplicity, we assume $a = a_+ = a_-$. From the individual energy contributions, the cumulative energy density for the nematic phase finally reads

$$F = F_{elast} + (F_{de} + F_{flex} + F_{int}) + F_{entropic} \tag{9}$$

Following the free energy form, we now proceed to minimize it with respect to the electrostatic potential as $\frac{\delta F}{\delta \psi} = 0$ that results in equation governing the distribution of potential and the ionic species within the liquid phase at equilibrium condition. The modified Poisson-type equation which couples the ionic and the potential distribution with director configuration reads

$$\varepsilon_0 (\varepsilon_a \cos^2(\theta) + \varepsilon_\perp) \frac{d^2 \psi}{dy^2} - \varepsilon_0 \varepsilon_a \left( E_x \cos(2\theta) + \sin(2\theta) \frac{d\psi}{dy} \right) \frac{d\theta}{dy} + \frac{1}{2}(e_1 - e_3) \sin(2\theta) \frac{d^2 \theta}{dy^2} \\ + (e_1 - e_3) \cos(2\theta) \left( \frac{d\theta}{dy} \right)^2 + ze(n_+ - n_-) = 0 \tag{10}$$

The corresponding electrochemical potential for the present system, that may be obtained as $\mu_\pm = \frac{\delta F}{\delta n_\pm}$ [59], is a gradient free quantity for equilibrium condition leading to the modified Boltzmann distribution as [62,63]

$$n_{\pm} = \frac{n_0 \exp(\mp ez\psi/k_B T)}{1 + \nu(\cosh(ez\psi/k_B T) - 1)} \tag{11}$$

In the above form $\nu = 2n_0 a^3$ denotes the steric factor and $n_0$, the number density of ions in the bulk reservoir. Substituting the Boltzmann distribution $n_{\pm}$ into Poisson equation, the modified Poisson-Boltzmann equation for the NLC phase is obtained. The cumulative charge within half the channel is equal and opposite to the charge density induced at the wall which provides us with the necessary boundary condition applicable to either of the walls. Consequently, the condition at the upper boundary reads

$$\sigma_w(y=1) = -\int_0^h \rho_e dy = \int_0^h \frac{2zen_\infty \sinh(ez\psi/k_B T)}{1 + \nu(\cosh(ez\psi/k_B T) - 1)} dy \tag{12}$$

where $\rho_e = e\sum_i z_i n_i = ez(n_+ - n_-)$ represents the net charge density. In the similar manner, the boundary condition for the lower plate can be derived.

From a fundamental rate of work hypothesis as proposed in the Leslie-Ericksen theory of nematodynamics, we obtain two governing equations, one for the sub-layer adjacent to each surface and one applicable within the bulk, for the angular momentum balance of the director [10] involving the cumulative effects of elastic, dielectric, flexoelectric energies and the fluid flow. The governing form valid within each of the two thin surface sub-layers; i.e. $-h \leq y \leq -(h-\delta)$ and $(h-\delta) \leq y \leq h$ reads

$$2K^* \frac{d^4\theta}{dy^4} - \left(K_1 \sin^2(\theta) + K_3 \cos^2(\theta)\right)\frac{d^2\theta}{dy^2} - (K_1 - K_3)\sin(\theta)\cos(\theta)\left(\frac{d\theta}{dy}\right)^2 - \left(\alpha_3 \sin^2(\theta) - \alpha_2 \cos^2(\theta)\right)\frac{du}{dy}$$
$$-\varepsilon_0 \varepsilon_a \left[\frac{1}{2}\sin(2\theta)\left(E_x^2 - \left(\frac{d\psi}{dy}\right)^2\right) - E_x \frac{d\psi}{dy}\cos(2\theta)\right] + \frac{1}{2}(e_1 - e_3)\sin(2\theta)\frac{d^2\psi}{dy^2} = 0 \tag{13}$$

while the form governing the bulk director dynamics i.e. $-(h-\delta) \leq y \leq (h-\delta)$ is given by

$$\left(K_1\sin^2(\theta)+K_3\cos^2(\theta)\right)\frac{d^2\theta}{dy^2}+(K_1-K_3)\sin(\theta)\cos(\theta)\left(\frac{d\theta}{dy}\right)^2+\left(\alpha_3\sin^2(\theta)-\alpha_2\cos^2(\theta)\right)\frac{du}{dy}$$
$$+\varepsilon_0\varepsilon_a\left[\frac{1}{2}\sin(2\theta)\left(E_x^2-\left(\frac{d\psi}{dy}\right)^2\right)-E_x\frac{d\psi}{dy}\cos(2\theta)\right]-\frac{1}{2}(e_1-e_3)\sin(2\theta)\frac{d^2\psi}{dy^2}=0 \quad (14)$$

Here we have considered the definitions of the director **n** and velocity **V** as given above. The boundary condition, as found by employing the variation of the total energy (equation (9)) at the boundary surface, yields four non-linear conditions that reads

$$K^*\theta'' - K_{13}\sin(2\theta) = 0 \quad \text{at } y=-h \qquad (15)$$

$$K^*\theta'' - K_{13}\sin(2\theta) = 0 \quad \text{at } y=-h \qquad (16)$$

$$K^*\theta''' - \left(K_{11}\sin^2(\theta)+K_{33}\cos^2(\theta)-2K_{13}\cos(2\theta)\right)\theta' - \left(e_1\sin^2(\theta)+e_3\cos^2(\theta)\right)E_x$$
$$-\frac{1}{2}(e_1-e_3)\sin(2\theta)E_y + \frac{W_S}{2}\sin(2(\theta-\theta_p))=0 \qquad \text{at } y=-h \quad (17)$$

and

$$K^*\theta''' - \left(K_{11}\sin^2(\theta)+K_{33}\cos^2(\theta)-2K_{13}\cos(2\theta)\right)\theta' - \left(e_1\sin^2(\theta)+e_3\cos^2(\theta)\right)E_x$$
$$-\frac{1}{2}(e_1-e_3)\sin(2\theta)E_y - \frac{W_S}{2}\sin(2(\theta-\theta_p))=0 \qquad \text{at } y=h \quad (18)$$

The sequence of equations for the potential distribution and director orientation must be closed by the balance of linear momentum governing the fluid flow velocity to determine the electroosmotic flow conditions for a nematic LC. Towards this, we employ the Leslie-Ericksen theory where the set of equations governing the linear momentum balance [9,10]. For the present study, an electroosmotic body force density $f_{eo}$ gets induced, where $f_{eo}=-\left(c_+\nabla\mu_+ + c_-\nabla\mu_-\right)$, that actuates the flow through the narrow conduit. The governing equation for the steady, electroosmotically driven flow of a nematic fluid through a narrow confined cell reduces to

$$\frac{d}{dy}\left(\eta(\theta)\frac{du}{dy}\right) = -\rho_e E_x \qquad (19)$$

Here the classical no-slip boundary condition is imposed at both the walls while $\rho_e E_x$ gives the electroosmotic body force density. The position-dependent apparent nematic viscosity is a function of the director alignment, that reads $\eta(\theta) = \eta_1 \sin^2 \theta + \eta_2 \cos^2 \theta + \eta_{12} \sin^2 \theta \cos^2 \theta$ where the viscosity parameters $\eta_1, \eta_2$ and $\eta_{12}$ are known as the Miesowicz viscosities, that, in turn, is related to the Leslie viscosities by the following relations: $\eta_1 = \frac{\alpha_3 + \alpha_4 + \alpha_6}{2}$, $\eta_2 = \frac{-\alpha_2 + \alpha_4 + \alpha_5}{2}$ and $\eta_{12} = \alpha_1$ [10]. Before proceeding to solving the electroosmotic flow of the nematic crystals, we proceed to derive a dimensionless set for the above governing equations and the corresponding non-linear boundary conditions resulting in a more general representation of the flow characteristics.

Next we proceed to adopt a suitable non-dimensionalization scheme, to obtain the dimensionless forms of the governing equations and boundary conditions, as follows: $\bar{y} = y/h$, $\bar{\psi} = ze\psi/k_B T$, $\bar{u} = u/u_{ref}$, $\bar{E}_x = E_x/E_{x,ref}$, $\bar{E}_y = E_y/E_{y,ref}$. Applying the small deformation limit $(\theta \to 0)$, we linearize the governing equations within the surface sub-layer and the corresponding boundary conditions while the governing equations beyond this region are solved in their usual forms [38]. Under these considerations the set of equations get reduced to the following forms:

*Dimensionless modified Poisson-Boltzmann equation:*

The dimensionless form for equation(11) for the thin near-surface region and the bulk is given as

$$\left.\begin{array}{l}
\text{within the surface transition layer i.e. } -h \leq y \leq -(h-\delta) \text{ and } (h-\delta) \leq y \leq h \\
\left(1 + \frac{\varepsilon_a}{\varepsilon_\perp}\right)\frac{d^2\bar{\psi}}{d\bar{y}^2} - \frac{\bar{E}_x}{p_1}\frac{d\theta}{d\bar{y}} - \frac{\sinh(\bar{\psi})}{\left(1+v(\cosh(\bar{\psi})-1)\right)\bar{\lambda}^2} = 0 \qquad (a) \\
\text{and} \\
\text{in the bulk region i.e. } -(h-\delta) \leq y \leq (h-\delta) \\
\left(1 + \frac{\varepsilon_a}{\varepsilon_\perp}\cos^2(\theta)\right)\frac{d^2\bar{\psi}}{d\bar{y}^2} - \left(\frac{\bar{E}_x \cos(2\theta)}{p_1} + \sin(2\theta)\frac{d\bar{\psi}}{d\bar{y}}\right)\frac{d\theta}{d\bar{y}} \\
+ A_4\left\{\sin(2\theta)\frac{d^2\theta}{d\bar{y}^2} + 2\cos(2\theta)\left(\frac{d\theta}{d\bar{y}}\right)^2\right\} - \frac{\sinh(\bar{\psi})}{\left(1+v(\cosh(\bar{\psi})-1)\right)\bar{\lambda}^2} = 0 \qquad (b)
\end{array}\right\} \quad (20)$$

$$\bar{\sigma}_w = \int_0^1 \frac{\sinh(\bar{\psi})}{\left(1+v(\cosh(\bar{\psi})-1)\right)\bar{\lambda}^2} d\bar{y} \qquad (21)$$

Here $E_{x,ref}$ scale is considered in the order of Freedericksz transition field $E_{c1}$ which is defined as the threshold electric field above which deformations in the nematic director is observed [10,64] while $p_1 = \dfrac{k_B T}{E_{c1} h z e}$ and $A_4 = \dfrac{z e (e_1 - e_3)}{2(\varepsilon_0 \varepsilon_\perp) k_B T}$. Also $\bar{\sigma}_w = \dfrac{z e h \sigma}{\varepsilon_0 \varepsilon_\perp k_B T}$ denotes the dimensionless surface charge density and $\bar{\lambda} = \dfrac{\lambda}{h} = \sqrt{\dfrac{\varepsilon_0 \varepsilon_\perp k_B T}{2 z^2 e^2 n_0 h^2}}$; $\lambda$ being the dimensional Debye screening length. It is to be noted that here the linearization is done only with respect to orientation angle $\theta$ but the frequently used Debye–Hückel linearization [15], which is valid for small electrostatic potential range, is not employed. Thus in terms of electrostatic potential $\psi$, the results of the present study will be comprehensive and general.

*Dimensionless form of angular momentum balance equation*:

The corresponding dimensionless form for the linearized equation(13) and the bulk equation(14) governing the angular momentum of the NLC phase reads

$$\left\{\begin{array}{l} \text{within the surface transition layer i.e. } -1 \leq \bar{y} \leq -(1-\bar{\delta}) \text{ and } (1-\bar{\delta}) \leq \bar{y} \leq 1 \\[4pt] b^2 \dfrac{d^4 \theta}{d\bar{y}^4} - \kappa \dfrac{d^2 \theta}{d\bar{y}^2} + m \bar{\alpha}_2 \dfrac{d\bar{u}}{d\bar{y}} - q\left[(\bar{E}_x^2 - p^2 \cdot \bar{E}_y^2) 2\theta + 2 \cdot p \cdot \bar{E}_x \cdot \bar{E}_y\right] - 2w \cdot \theta \dfrac{d\bar{E}_y(\bar{y})}{d\bar{y}} = 0 \quad (a) \\[6pt] \text{and} \\[4pt] \text{in the bulk region i.e. } -(1-\bar{\delta}) \leq y \leq (1-\bar{\delta}) \\[4pt] \left(\sin^2(\theta) + \kappa \cos^2(\theta)\right) \dfrac{d^2 \theta}{d\bar{y}^2} + (1-\kappa)\sin(\theta)\cos(\theta)\left(\dfrac{d\theta}{d\bar{y}}\right)^2 + m\left(\bar{\alpha}_3 \sin^2(\theta) - \bar{\alpha}_2 \cos^2(\theta)\right) \dfrac{d\bar{u}}{d\bar{y}} \\[4pt] + q\left[(\bar{E}_x^2 - p^2 \cdot \bar{E}_y^2)\sin(2\theta) + 2 \cdot p \cdot \bar{E}_x \cdot \bar{E}_y \cos(2\theta)\right] + w \cdot \sin(2\theta) \dfrac{d\bar{E}_y(\bar{y})}{d\bar{y}} = 0 \quad (b) \end{array}\right. \quad (22)$$

The various dimensionless parameters introduced in the above equation are defined as $\kappa = K_{33}/K_{11}$, $\bar{\alpha}_3 = \alpha_3/\eta_{ref}$, $E_{y,ref} = \dfrac{E_y}{\sigma_w / \varepsilon_0 \bar{\varepsilon}}$, $\bar{\alpha}_2 = \alpha_2/\eta_{ref}$, $q = \dfrac{\varepsilon_0 \varepsilon_a E_{x,ref}^2 h^2}{2 K_1}$, $m = \dfrac{u_{ref} h \eta_{ref}}{K_1}$, $p = \dfrac{\sigma_w}{\varepsilon_0 \bar{\varepsilon} E_{x,ref}}$ and $w = \dfrac{e_1 - e_3}{2 K_1}\left(\dfrac{\sigma h}{\varepsilon_0 \bar{\varepsilon}}\right)$ with $\bar{\varepsilon}$ being the average dielectric constant defined as $\bar{\varepsilon} = (\varepsilon_\parallel + 2\varepsilon_\perp)/3$. Here $b$ is a dimensionless characteristic interaction length, defined as

$b = \frac{1}{h}\sqrt{\frac{2K^*}{K_{11}}}$. The reference viscosity has been chosen as $\eta_{ref} = \alpha_4/2$ which is the Newtonian counterpart of the NLC viscosity as can be deduced from the deviatoric stress equation. while the velocity reference $u_{ref}$ can now finally be obtained from the equation mentioned below.

*Dimensionless form of linear momentum balance equation:*

Corresponding to the linear momentum balance equation(19) for the NLC fluid, we obtain its dimensionless form employing the aforementioned dimensional parameters that reads

$$\frac{d}{d\bar{y}}\left(\bar{\eta}(\theta)\frac{d\bar{u}}{d\bar{y}}\right) = \sinh(\bar{\psi})\bar{E}_x \qquad (23)$$

where the dimensionless viscosity function for the two regions are given as

$$\left.\begin{array}{l} \text{within the surface transition layer i.e. } -1 \leq \bar{y} \leq -(1-\bar{\delta}) \text{ and } (1-\bar{\delta}) \leq \bar{y} \leq 1: \\ \bar{\eta}(\theta) = \frac{\eta(\theta)}{\eta_{ref}} = (\eta_2/\eta_{ref}) \qquad (a) \\ \text{and} \\ \text{in the bulk region i.e. } -(1-\bar{\delta}) \leq y \leq (1-\bar{\delta}): \\ \bar{\eta}(\theta) = \frac{\eta(\theta)}{\eta_{ref}} = \sin^2\theta + (\eta_2/\eta_{ref})\cos^2\theta + (\eta_{12}/\eta_{ref})\sin^2\theta\cos^2\theta \qquad (b) \end{array}\right\} \qquad (24)$$

The velocity scale $u_{ref} = \frac{2zen_0 E_{x,ref} h^2}{\eta_{ref}}$ is used while reaching at the dimensionless equation (23). Owing to the linearized form of the governing equations very close to the boundary, we restrict our solutions to the case where the tilt angle of the director at the boundary $\theta_S$ remains close to the pretilt angle $\theta_p$. It must be noted that the highly non-linear set of governing equations couples the flow velocity and the director configuration with the potential distribution, a fact which is explicitly absent in case of electroosmotic flows of Newtonian fluids. In what follows, we consider a homeotropic alignment with pretilt angle equal to zero and obtain the director configuration, potential distribution and velocity profile for the NLC electroosmotic flow. An intriguing aspect we further put forward in this study is the director tilt at the boundary which depends non-linearly with the surface contributions and second-order elastic energies.

**Results and Discussion**

In the present section, we demonstrate the variation of the elastic and electrostatic surface energies on the director orientation and flow characteristics for an electroosmotic flow within the NLC cell. For a representative case, we have selected the nematic 5CB (4-Cyano-4'-pentylbiphenyl) for our calculation whose properties are detailed in **Table 1**. The controllable dimensionless parameters are chosen carefully keeping in view of the corresponding dimensional parameters involved. We have assumed that a selective adsorption of negative charges is taking place at the limiting surfaces. The induced surface charge density is varied between $10^{-4} - 10^{-2}$ Cm$^{-2}$ while a bulk concentration of ionic impurities is considered in order of $10^{-3}$ mM [46,65]. These result in a dimensionless Debye screening length range of $10^{-2} - 10^{-1}$ and the dimensionless surface charge density in the range of $\bar{\sigma}_w \sim 10^1 - 10^3$ if the channel half thickness is varied as $h \sim 1 - 10 \mu\text{m}$. The characteristic length $\sqrt{2K^*/K_{11}}$ varies in the order of molecular interaction (typically $20 \overset{0}{\text{A}}$) [38,66] giving rise to a dimensionless characteristic interaction length $\left( b = \frac{1}{h}\sqrt{\frac{2K^*}{K_{11}}} \right)$ in the range of $10^{-3} - 10^{-2}$. The surface anchoring energy parameter $\bar{\gamma} = \frac{W_S h}{2K_{11}}$ and the easy axis direction $\theta_p$ not only depend on the substrate with which it is covered but also on the surface alignment technique. In the present study we consider the easy direction perpendicular to the substrate ($\theta_p = 0$) i.e. the homeotropic alignment. In practical applications this situation is often realized with surface alignment techniques like stacking of amphiphilic molecules, oblique evaporation of SiO [67], deposition of monolayer lipid membrane on SiO$_2$ substrates [26] or topographical patterning of polymer films [68]. Following several experimental observations it is found that $W_S$ remains in the range of $10^{-6}$ to $5 \times 10^{-5}$ J/m$^2$ [22,25,69]. Thereafter, using aforementioned values of $h$ and $K_{11}$, the range of dimensionless anchoring energy parameter $\bar{\gamma} = \frac{W_S h}{2K_{11}}$ can be obtained as $\bar{\gamma} \sim 0.01 - 5$. The mixed splay-bend elastic constant relates to the splay elastic constant as $K_{13} = -0.2 K_{11}$, which is experimentally observed by Lavrentovich and Pergamenshchik [22]. In the absence of exact experimental data on the relation between surface transition layer thickness $\delta$ and the characteristic length $b$, the dimensionless thickness $\bar{\delta}$ will be considered to be twice the characteristic length $b$. Since we will consider the parametric variation of $b$ in the following results, the variation of $\bar{\delta}$ will also get inherently incorporated. In this section, we sort for the influence of the surface effects from $b, \bar{\gamma}, \bar{E}_x, \bar{\lambda}$ and $\bar{\sigma}_w$ on the resulting charge distribution, director orientation and the nature of the electroosmotic flow velocity. Unless otherwise mentioned the base values of these parameters are chosen as $\bar{b} = 0.01$, $\bar{\gamma} = 1$, $\bar{E}_x = 2$, $\bar{\lambda} = 0.1$ and $\bar{\sigma}_w = -2000$.

| Property | Property Value | Unit | Property | Property Value | Unit |
|---|---|---|---|---|---|
| Splay elastic constant | $K_{11} = 6.2$ | pN | | $\alpha_1 = -0.0060$ | |
| Bend elastic constant | $K_{33} = 8.2$ | pN | Leslie viscosity coefficients [10] | $\alpha_2 = -0.0812$ | |
| Dielectric permittivity (relative) | $\varepsilon_\parallel = 18.5$ and $\varepsilon_\perp = 7$ | — | | $\alpha_3 = -0.0036$ | Pa-s |
| | | | | $\alpha_4 = 0.0652$ | |
| | | | | $\alpha_5 = 0.0640$ | |
| Flexoelectric coefficients [65] | $e_1 = -25$ and $e_3 = -8.5$ | pC/m | | $\alpha_6 = -0.0208$ | |

**Table 1:** Details the symbols, magnitudes and units of the 5CB nematic properties used for the present study.

Flows of NLC fluids may be characterized by topological defects which results in singularity of director definition [9,10]. The existence of topological defects in NLC flows greatly depends on the channel dimensions and flow rates. [70] A dimensionless quantity, namely the Ericksen number $\left( Er = \dfrac{\eta_c u_c L_c}{K_c} \right)$ which signifies the relative importance of the viscous torque with respect the viscous torque, is often defined in this context. For the present case the characteristic viscosity $(\eta_c)$ and the characteristic elastic constant $(K_c)$ can be taken as $\eta_c = (\eta_1 + \eta_2)/2$ and $K_c = (K_{11} + K_{22} + K_{33})/3$, respectively [71] while the characteristic velocity is the average flow velocity $(u_{av})$ and characteristic length $(L_c)$ is the channel half height $(h)$. Both experimental [11,70] as well as theoretical studies [72–74] exist in literature which show that the topological defects become significant when the $Er$ is very high. On the other hand, we find that the actual maximum value of $Er$ for the present problem falls within the order of ~10; although in most of the cases it remains well below or around unity. Thus, for the present set of parameters considered, we can safely consider the flow to be "elastically" laminar with the absence of topological defects and the present formulation, following the LE formalism, remains valid.

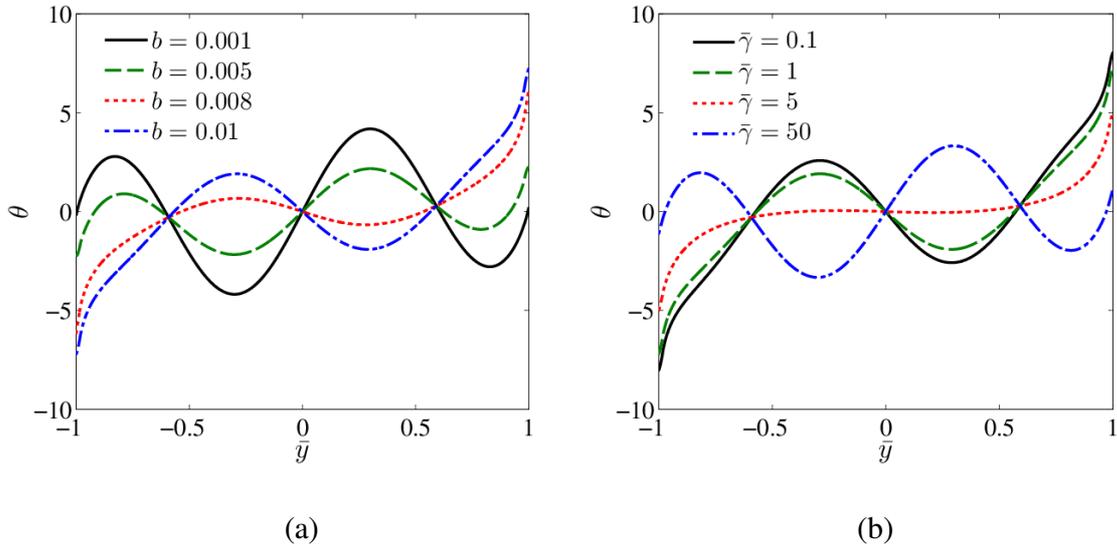

(a)        (b)

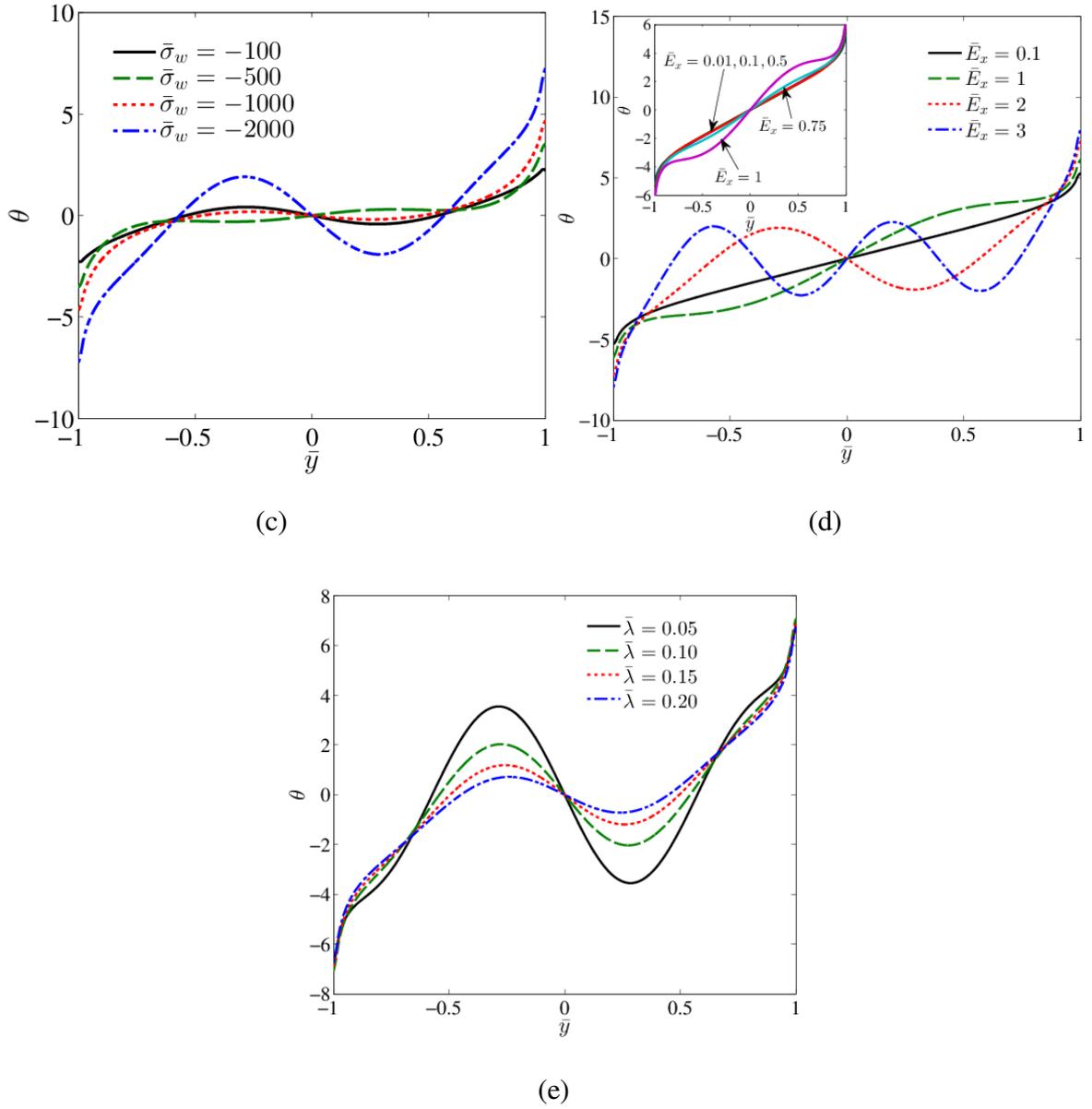

**Fig 2:** Depicts the variation of the director alignment profile $\theta$ as a function of the channel transverse direction $\bar{y}$ for different values of dimensionless a) second-order elastic constant, b) surface anchoring strength, c) surface charge density, d) axial applied field $\bar{E}_x$ and e) Debye length.

Figure 2 depicts the variation of the surface director orientation and the director alignment profile across the channel with different dimensionless controlling parameters. Before exploring the effects of the individual parameters, we note few general important characteristics of the director distribution. We observe a periodic pattern of the director alignment, the amplitude of which varies with different flow conditions. It is further seen that the director distortion at the surface is small and remains close to the homeotropic alignment, a result consistent with the assumption for inclusion of the second-order elasticity. The aspect of anti-symmetry with the director alignment [66] has also been captured. With these general considerations, we proceed to reflect on the influence of individual parameters on the director distortion. In figure 2a it is observed that with increased second-order elastic coefficient, there occurs a sharper surface distortion of the director. These characteristic director orientations confined near the surface depletion region lead to a varied director configuration in the across the channel, which implicitly reflects the impact of the surface phenomena into the bulk. Figure 2b depicts the effect of the surface anchoring strength on the director alignment. It is clearly evident that as the anchoring strength $\bar{\gamma}$ is enhanced, the boundary asymptotically exhibit a strong anchoring behavior, thereby, imposing the easy axis alignment on the director at the boundaries, which in our case is the homeotropic alignment ($\theta \to 0$). Interestingly, a subtle observation is that with higher anchoring strength, the effect of second-order surface elasticity sharply reduces. This is attributed to the very fact that such elasticity effects become negligible or remains absent in cases of strong anchoring boundary situations. It is also counter-intuitive to observe that the anchoring strength, though only appears in the mathematical equations only through the boundary conditions, it shows an evident effect on the bulk director distortion behavior. For a intermediate value of the said parameter, the periodic behavior is absent while it gets prominent for both high and low limits of $\bar{\gamma}$. Figure 2c depicts the director configuration across the channel for different values of induced surface charge density. The induced charge has an intrinsic effect on the director deformation at the surface due to a coupled effect of the transverse electric field induced aligning and flow induced aligning. With an increase in surface charge, the transverse field tends to orient the director in homeotropic alignment while the increased flow (as seen in figure 4b) tends to shift such orientation, resulting in the configuration as seen above. Figure 2d describes the director configuration for different values of the applied axial electric field $\bar{E}_x$. It is seen that as the applied field is increased, the director tries to orient itself along the field near the boundaries, thus deviating further from the perfect homeotropic limit. Also, with higher applied field, the frequency of the periodic configuration of the director increases. Such a periodic configuration has been experimentally observed in non-linear Electroosmosis flow of NLCs [12].

In the inset of figure 2d it is shown that the application of an applied axial field $(\bar{E}_x)$ will only influence the equilibrium director configuration if $(\bar{E}_x)$ crosses a threshold value $(\bar{E}_x \geq \bar{E}_{x,c})$. Till this threshold value of the electric field is reached (e.g. $\bar{E}_{x,c} \sim 0.7$ for the presented case), the directors assume a configuration corresponding to the non-flow condition. This phenomena has a resemblance to the so called *Freedericksz Transition*, widely introduced in LC literatures [10].

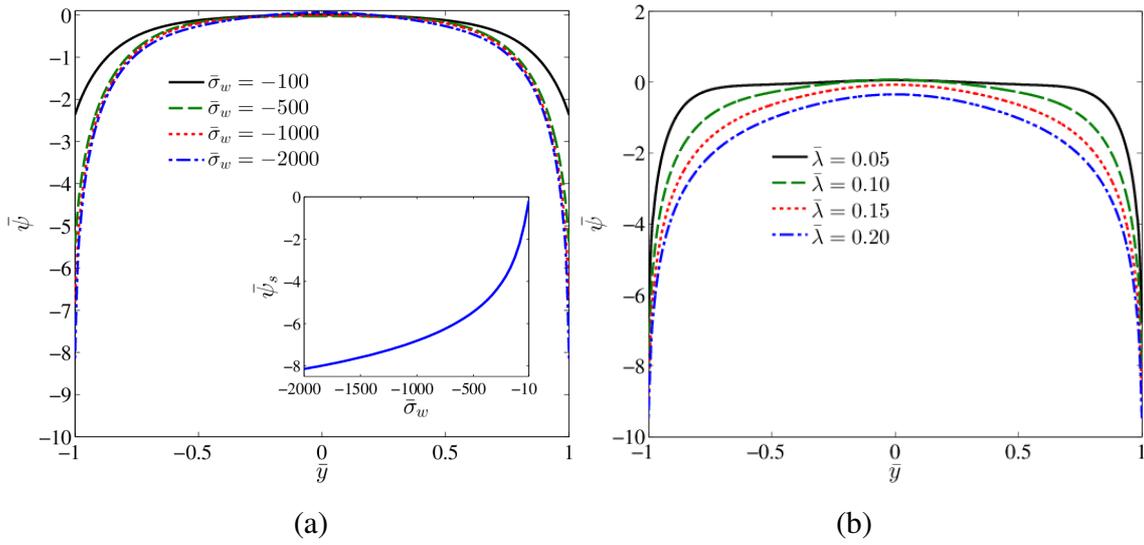

(a)       (b)

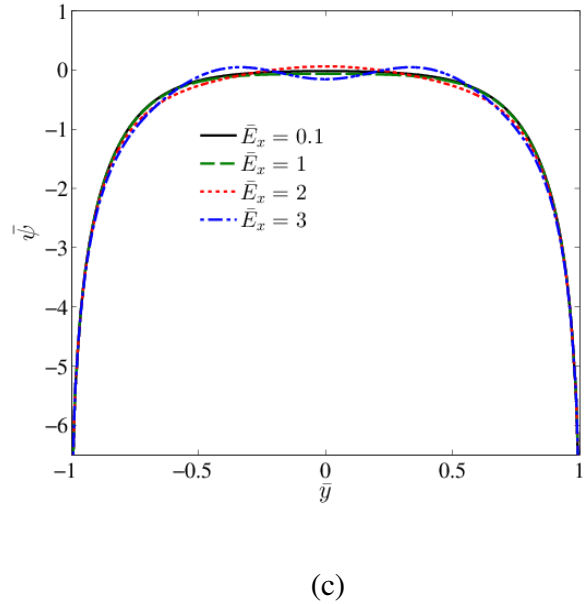

(c)

**Fig 3:** Depicts the variation of the potential distribution $\bar{\psi}$ as a function of the channel transverse direction $\bar{y}$ for different values of dimensionless a) surface charge density, b) Debye length and c) axial applied field $\bar{E}_x$.

Figure 3 illustrates the dimensionless equilibrium potential distribution due to the induced EDL across the channel section. Due to the assumption of a induced negative charge at the substrate surfaces, a potential distribution with negative potential is observed. The resultant induced field not only affects the flow but also influences the director configuration which, in turn, affects the flow rheology. In figure 3a, we note the variation of the potential distribution for different values of the induced surface charge density. Higher surface charge implies a higher potential magnitude, and thereby, a stronger transverse electric field. A stronger electric field implies that the field attempts to orient the director along its direction besides inducing a higher body force for the flow actuation. A coupled effect results in the director configuration as observed in figure 2c. The surface potential, however, does not linearly increase with the surface charge as seen in the inset. This leads to a non-linear variation of the flow velocity with increase in $\bar{\sigma}_w$. Figure 3b relates the potential distribution for different values of the dimensionless Debye length $\bar{\lambda}$. The factor $\bar{\lambda}$ signifies the apparent penetration of Debye length into the channel centerline. Consequently, the electrical double layer and the induced transverse field dominate across a larger span of the channel the result of which is manifested in a larger flow velocity as will be seen later. Figure 3c illustrates the potential distribution variation across the channel due to the applied electric field. It must be noted here that for steady electroosmotic flows of aqueous electrolyte through slit geometries, the potential distribution remains unaffected by the applied axial field. However, for NLC medium, the field has a direct effect on the director distribution which intrinsically modifies the potential distribution across the channel. This intercoupling of field dominated director configuration and charge distribution is clearly manifested especially for higher electric fields, as evident from figure 3c.

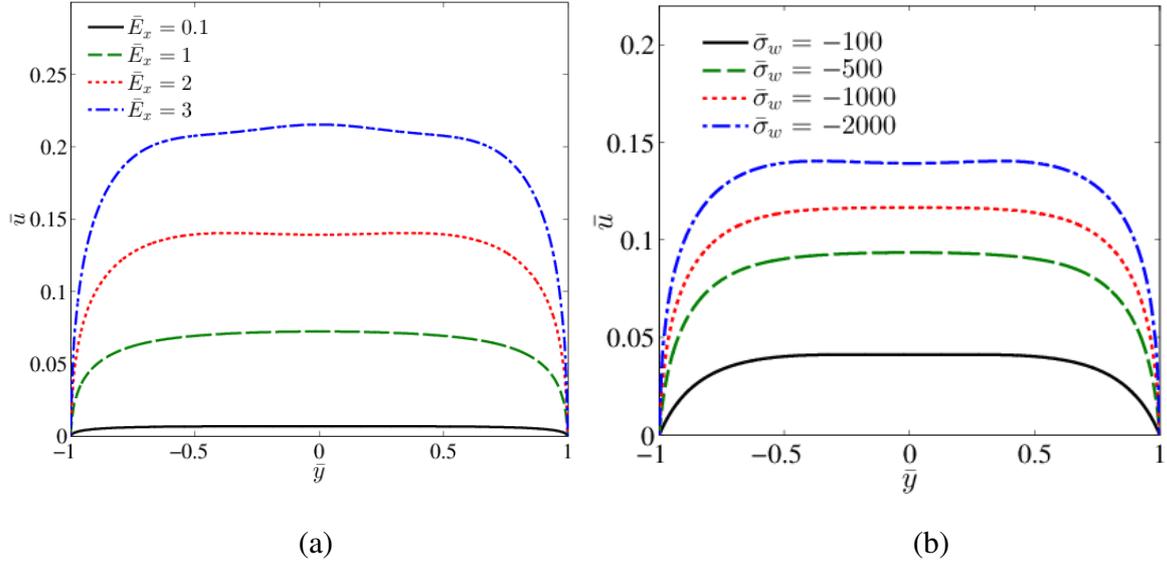

(a)  (b)

**Fig 4:** Depicts the variation of the velocity field profile $\bar{u}$ as a function of the channel transverse direction $\bar{y}$ for different values of dimensionless a) axial applied field $\bar{E}_x$ and b) surface charge density $\bar{\sigma}_w$.

Figure 4 illustrates the velocity profile characteristics of electroosmotic flow in NLC fluids for different values of a) axial field $\bar{E}_x$ and b) surface charge density $\bar{\sigma}_w$. In both the cases, it can be seen that the velocity profile, in similarity with electroosmotic flows for Newtonian fluids, follows a region of high velocity gradient near the wall and an apparent plug region at the bulk. With an increase in either the axial field or the surface charge density, the flow velocity gets augmented. This is directly accredited to the enhanced electroosmotic body force due to a rise in either the actuating field or the induced surface charge. The velocity profile for an electroosmotic flow of liquid crystals experiences drastic variations depending on various factors related to surface anchoring, genesis of EDL and the mode of electroosmotic flow. In past literature, it has been experimentally observed that an application of an external electrical field results in induced ion generation leading to the formation of a double layer. The interaction of this double layer with the applied field, in turn, drives a flow where the velocity profile depicts regions of opposite flow patterns. Such flow actuation mechanism belongs to the category of non-linear electroosmotic flow. However, in sharp contrast to the above situation, we consider the generation of EDL at the fluid-substrate interface is independent of the applied field. Further, with the consideration of an weak anchoring condition at the boundaries, the sharp changes in the director configuration remains suppressed which translates into a velocity profile similar to that of electroosmosis in Newtonian medium. It must nevertheless be appreciated that with higher applied field, the amplitude of the periodic pattern of the director alignment, which is also observed in non-linear electroosmotic flows, gets enhanced and induce slight characteristic undulations in the resulting velocity profile. Such undulation pattern in the velocity field gets further augmented with stronger anchoring strength or strong anchoring condition.

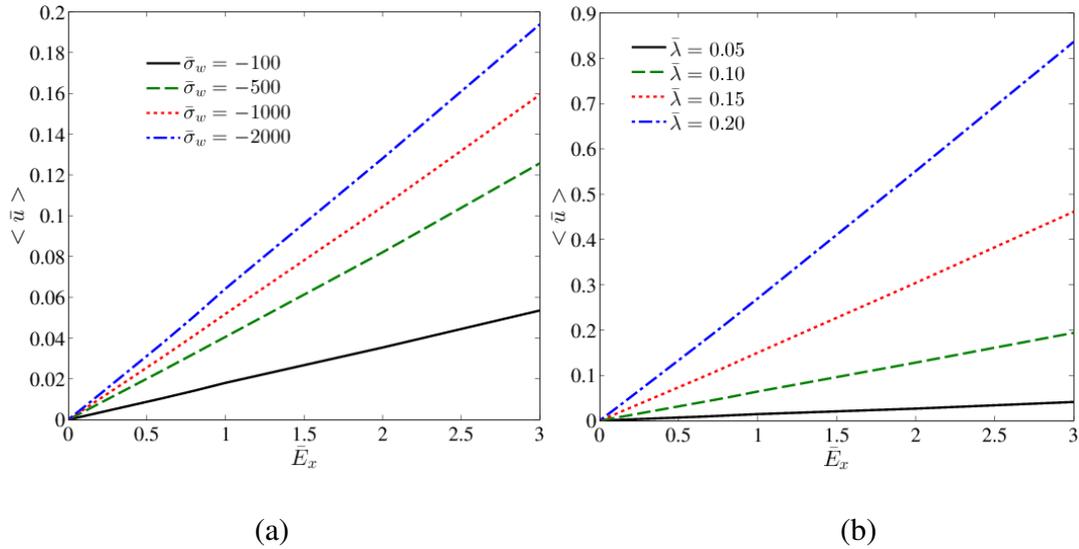

(a) (b)

**Fig 5:** Depicts the variation of the average velocity as a function of the axial field $\bar{E}_x$ for different values of the a) dimensionless surface charge density and b) dimensionless Debye screening length $\bar{\lambda}$.

Figure 5 depicts the average flow velocity $<\bar{u}>$ for the electroosmosis of NLC for different values of $\bar{\sigma}_w$ and $\bar{\lambda}$. A direct conclusion from the average flow velocity variation clearly suggests the flow characteristics belong to linear electroosmosis system wherein the velocity varies linearly with the electric field. Figure 5a illustrates that the average velocity gradually increases with an increase in the surface charge density. This is attributed to the corresponding increase of the transverse field with higher surface charge that enhances the effective flow body force, and thereby, the flow average velocity. However, since the increase of the surface potential, and thereby, the transverse field does not vary linearly, as seen in the inset of figure 3c, the increase in the average velocity rate diminishes with $\bar{\sigma}_w$. Figure 5b also shows that the average velocity magnitude increases with increase in $\bar{\lambda}$. As the value of $\bar{\lambda}$ increase, the penetration of Debye length towards the channel centerline increases as shown in figure 3b. This means a larger section of the nematic fluid experiences the electroosmotic body force that results in a higher flowrate.

**Conclusion**

Electroosmosis of nematic liquid crystals fluids have serious technological and academic implications in the purview of electrokinetic transport phenomenon. Electrically actuated flows of complex NLC fluids through micro-confinement with surface dominated characteristics is the focus of the present study. Governing formulation of the problem is devised based on fundamental free energy considerations taking into account the intricate anisotropic dielectric and viscous features of the NLC medium. Since the fluid flows through a micro-confinement, surface effects become prominent. Due to proper characterization of such surface influences, second order elastic theory and second order elastic weak anchoring energies have been considered to model the director configuration with the assumption of small deformations near the surface. The EDL is modeled with a modified Poisson-Boltzmann equation considering excluded volume effects while the equations are closed with the LE theory governing the electroosmotic flow velocity profile. Interesting optical periodic stripes are observed for director configuration for higher flow rates and electric fields. The average velocity is seen to increase linearly with the electric field indicating the flow is in purview of linear electroosmosis.